# On the impact of correlation information on the orientation parameters between celestial reference frame realizations

Yulia Sokolova[1], Zinovy Malkin[1,2]

[1]*Pulkovo Observatory, St. Petersburg, Russia*
[2]*St. Petersburg State University, St. Petersburg, Russia*

In this study, we compared results of determination of the orientation angles between celestial reference frames realized by radio source position catalogues using three methods of accounting for correlation information: using the position errors only, using additionally the correlations between the right ascension and declination (RA/DE correlations) reported in radio source position catalogues published in the IERS format, and using the full covariance matrix. The computations were performed with nine catalogues computed at eight analysis centres. Our analysis has shown that using the RA/DE correlations only slightly influences the computed rotational angles, whereas using the full correlation matrices leads to substantial change in the orientation parameters between the compared catalogues.

## 1 Introduction

Catalogues of radio source positions derived from VLBI observations are used by the International Astronomical Union (IAU) to establish the International Celestial Reference Frame (ICRF) since 1998 (Ma et al. 1998; Feissel and Mignard 1998; Ma et al. 2009). In addition to the official ICRF catalogues International VLBI Service for Geodesy and Astrometry (IVS) analysis centres routinely publish radio source catalogues (RSC), hereafter referred to as individual catalogues, which, generally speaking, represent independent celestial reference frame (CRF) realizations. Comparison of these catalogues both among themselves and with the ICRF is important for quality assessment of the ICRF. The primary interest is to investigate the mutual orientation of these individual systems, which can be represented by rotation around three Cartesian axes by the angles A1, A2, A3. In this study we investigate the impact of correlations between radio source positions on determination of these angles, which is an extension of the first study on the subject of Jacobs et al. (2010).

## 2 Method used

For transformation of a radio source vector (X, Y, Z) from one system to another we can write for small rotation angles:

$$\begin{pmatrix} X_1 \\ Y_1 \\ Z_1 \end{pmatrix} = \begin{pmatrix} 1 & A_3 & -A_2 \\ -A_3 & 1 & A_1 \\ A_2 & -A_1 & 1 \end{pmatrix} \begin{pmatrix} X_2 \\ Y_2 \\ Z_2 \end{pmatrix}. \qquad (1)$$

Then, taking into account that

$$\begin{pmatrix} X \\ Y \\ Z \end{pmatrix} = \begin{pmatrix} \cos\alpha \cos\delta \\ \sin\alpha \cos\delta \\ \sin\delta \end{pmatrix}, \qquad (2)$$

where α is the right ascension and δ is the declination of the source, we find for the difference of radio source coordinates in the compared catalogues $\Delta\alpha = \alpha_1 - \alpha_2$ and $\Delta\delta = \delta_1 - \delta_2$:

$$\Delta\alpha \cos\delta = A_1 \cos\alpha \sin\delta + A_2 \sin\alpha \sin\delta - A_3 \cos\delta,$$
$$\Delta\delta = -A_1 \sin\alpha + A_2 \cos\alpha. \quad (3)$$

The rotation angles $A_1$, $A_2$, and $A_3$ are obtained by applying the least squares method for the equations (3) on all sources or selected group of sources. In our investigation we used all common sources between compared catalogues.

This method was used by the International Earth Rotation and Reference Systems Service (IERS) for comparison and combination of individual radio source catalogues in the 1980s and 1990s, before adopting the official version of the ICRF (Arias et al. 1988). Although systematic differences between individual catalogues are much more complicated than represented by the simple rotation model (Sokolova and Malkin 2007), determining orientation of the frames is a fundamental part of the constructing ICRF, like it was done for ICRF and ICRF2 (Ma et al. 1998; Ma et al. 2009). In both cases the computation of the final reference frame was conducted in two steps. First, the catalogue was calculated as a result of the global VLBI solution in its own independent system, and in the second step the catalogue was transferred to the IERS95 system in the case of ICRF and to the ICRF system in the case of ICRF2 applying the rotation model on a set of defining sources.

In order to obtain parameters of the rotation model (3) the following system is solved by least squares:

$$\mathbf{Bx} + \mathbf{\varepsilon} = \mathbf{l}, \quad (4)$$

where **B** is the coefficient matrix of dimension $2n \times 3$, $\mathbf{x} = (A_1, A_2, A_3)'$ is the vector of unknowns, $\mathbf{\varepsilon}$ is the vector of errors with length $2n$, $\mathbf{l} = (\Delta\alpha_1 \cos\delta_1, \Delta\delta_1, ..., \Delta\alpha_n \cos\delta_n, \Delta\delta_n)'$ is the vector of length $2n$ of coordinate differences, $n$ is the number of sources used, and prime denotes the transpose of a matrix. Note that since we have two Eqs. (3) for each source, the dimension of corresponding arrays is $2n$.

The covariance matrix for coordinate differences is given by

$$\mathbf{Q} = \mathbf{Q}_1 + \mathbf{Q}_2, \quad (5)$$

where $\mathbf{Q}_1$ and $\mathbf{Q}_2$ are the covariance matrices of the compared catalogues of dimension $2n \times 2n$. Each of them may be either a diagonal matrix if only position uncertainties are used or a two-diagonal matrix if the correlations between the right ascension and declination estimates for each source (RA/DE correlations) are taken into account or a full matrix when available. Strictly speaking the Eq. (5) is correct if the catalogues are independent. But the investigation of source coordinate correlations between catalogues is a separate nontrivial task, which is not addressed here. A possible approach to its solution is discussed by Malkin (2013), but not all problems are solved yet.

Finally, the solution of the system (4) is

$$\mathbf{x} = (\mathbf{B}'\mathbf{Q}^{-1}\mathbf{B})^{-1}\mathbf{B}'\mathbf{Q}^{-1}\mathbf{l}. \quad (6)$$

Traditionally, the system (4) is solved with weights of condition equations (3) inversely proportional to the position uncertainties reported in the catalogues. In other words, a diagonal covariance matrix is used in the least square solution. However, Jacobs et al. (2010) showed that accounting for correlations between the source positions derived from VLBI global solution changes significantly the orientation parameters between CRF realizations if a microarcsecond level of accuracy is required. In their work, catalogues with diagonal covariance matrix and with full covariance matrix were used for an investigation of the impact of correlation information on the orientation angles.

Currently, the IVS analysis centres mostly provide their individual CRF solutions in a standard IERS format, where the RA/DE correlations for each source are reported together with radio source positions and other relevant information. The ICRF2 catalogue is also published in the same format. So, in this case we only can use a two-diagonal covariance matrix in the least square solution. This case not considered by Jacobs et al. (2010) is also important to investigate, since such a research has not been done yet.

Recently some IVS analysis centres have started producing results of global solutions in the SINEX format, where the full covariance matrix is presented. For our work, we used two such solutions provided by Vienna University of Technology, Austria and NASA Goddard Space Flight Center, USA. Thus we could undertake a detailed comparison of commonly used rotational alignment models with three methods of accounting for the covariance information: using the position errors only (diagonal covariance matrix), using additionally RA/DE correlations available for the catalogues in the IERS format (two-diagonal covariance matrix), and using the full covariance matrices from SINEX files.

## 3  Results of computations

Nine CRF solutions from eight IVS analysis centres have been used for our investigation (Table 1): AUS (Geoscience Australia ), BKG (Federal Agency for Cartography and Geodesy, Germany), CGS (Space Geodesy Centre, Italy), GSFC (NASA Goddard Space Flight Center, USA), VIE (Vienna University of Technology, Austria), OPA (Paris Observatory, France), and SHA (Shanghai Astronomical Observatory, China).

Table 1. CRF solutions used in this study

| Catalogue | Analysis centre | Software | Time span | Number of sources | Format |
|---|---|---|---|---|---|
| aus2012b | AUS | Occam | 1980 – 2012 | 2892 | ICRF |
| bkg2011a | BKG | Calc/Solve | 1984 – 2011 | 3214 | ICRF |
| cgs2012a | CGS | Calc/Solve | 1980 – 2011 | 842 | ICRF |
| gsf2011a | GSF | Calc/Solve | 1979 – 2011 | 1340 | SINEX |
| gsf2012a | GSF | Calc/Solve | 1979 – 2012 | 3708 | ICRF |
| vie2012a | VIE | VieVS | 1984 – 2011 | 860 | SINEX |
| opa2012a | OPA | Calc/Solve | 1979 – 2012 | 3482 | ICRF |
| sha2012b | SHA | Calc/Solve | 1979 – 2012 | 3470 | ICRF |
| usn2012a | USN | Calc/Solve | 1979 – 2012 | 793 | ICRF |

The results of computation of the orientation angles between seven catalogues published in the IERS format and ICRF2 are shown in Table 2. In this test, diagonal or two-diagonal covariance matrices $\mathbf{Q}_1$ and $\mathbf{Q}_2$ were used.

Table 2. Orientation parameters between individual CRF solutions and ICRF2 computed with diagonal (first line) and two-diagonal (second line) covariance matrices. The number of common sources between the individual catalogue and ICRF2 is given below the catalogue name. Unit: µas

| Catalogue and source number | A1 | A2 | A3 |
|---|---|---|---|
| aus2012b 936 | −26.5 ± 4.5<br>− 26.5 ± 4.5 | 2.1 ± 4.6<br>2.3 ± 4.5 | 1.7 ± 4.0<br>2.5 ± 3.9 |
| bkg2011a 936 | 25.6 ± 3.1<br>25.5 ± 3.0 | 17.0 ± 3.1<br>17.1 ± 3.1 | −10.5 ± 2.7<br>− 12.9 ± 2.6 |
| cgs2012a 795 | 11.9 ± 3.5<br>12.0 ± 3.5 | −1.1 ± 3.5<br>− 0.2 ± 5.5 | −13.0 ± 3.1<br>− 18.2 ± 3.0 |
| gsf2012a 936 | −0.9 ± 2.3<br>− 1.0 ± 2.3 | 6.0 ± 2.3<br>5.8 ± 2.3 | −4.9 ± 2.0<br>− 3.7 ± 2.0 |
| opa2012a 936 | −4.7 ± 2.1<br>− 5.0 ± 2.1 | 10.8 ± 2.1<br>10.9 ± 2.1 | −10.0 ± 1.8<br>− 10.5 ± 1.8 |
| sha2012b 936 | −4.4 ± 2.2<br>− 4.4 ± 2.1 | 2.3 ± 2.2<br>2.4 ± 2.2 | −5.2 ± 1.9<br>− 5.1 ± 1.9 |
| usn2012a 780 | −2.5 ± 2.6<br>− 2.9 ± 2.6 | 10.5 ± 2.6<br>10.7 ± 2.6 | −6.5 ± 2.3<br>− 5.6 ± 2.3 |

Table 3 shows the results of computation of the orientation angles between two catalogues with full covariance matrices and ICRF2. In this test, we used diagonal, two-diagonal, or full covariance matrices for GSF and VIE catalogues and diagonal or two-diagonal covariance matrix for ICRF2.

Table 3. Orientation parameters between individual CRF solutions and ICRF2 computed with diagonal (first line), two-diagonal (second line), and full (third line) covariance matrices. The number of common sources between the catalogue and ICRF2 is given below the catalogue name. Unit: µas

| Catalogue | A1 | A2 | A3 |
|---|---|---|---|
| gsf2011a 1171 | −9.2 ± 2.3<br>−9.3 ± 2.3<br>−4.4 ± 3.1 | 3.5 ± 2.3<br>3.7 ± 2.3<br>2.8 ± 3.1 | 2.2 ± 2.0<br>2.0 ± 1.9<br>0.3 ± 2.3 |
| vie2012b 856 | 13.3 ± 3.0<br>13.2 ± 3.0<br>2.6 ± 4.1 | 11.3 ± 3.0<br>11.5 ± 3.0<br>7.0 ± 4.3 | −1.1 ± 2.4<br>−0.8 ± 2.3<br>− 0.9 ± 2.4 |

Table 4 presents results of comparisons of two catalogues with full covariance matrices to each other. In this test we used all three modes of covariance matrices for both catalogues.

Table 4. Orientation parameters between gsfc2011a and vie2012b catalogues computed with diagonal (first line), two-diagonal (second line), and full (third line) covariance matrices. The number of common sources between the catalogues is given below the catalogue name. Unit: µas

| Catalogues | A1 | A2 | A3 |
|---|---|---|---|
| gsfc2011a vie2012b 854 | −24.0 ± 2.1<br>−23.9 ± 2.1<br>− 0.6 ± 2.5 | −5.4 ± 2.2<br>−5.2 ± 2.2<br>3.4 ± 3.5 | 6.0 ± 1.1<br>5.7 ± 1.1<br>0.7 ± 1.2 |

Obtained results have shown that the off-diagonal elements of the catalogue covariance matrix have large impact on the results of computation of the orientation parameters. Figure 1 gives an impression of how large these correlations can be. Although generally the off-diagonal correlations are small, some can exceed 0.9.

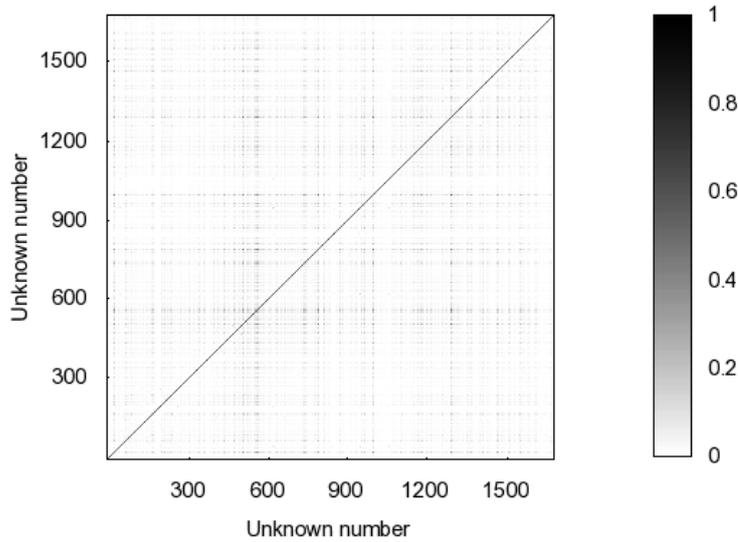

Fig. 1. Correlation matrix of the catalogue gsfc2011a for sources used in the computations presented in Table 4. For better visibility, the absolute values of correlations are depicted. The odd unknown numbers correspond to the right ascension (RA) of the sources sorted by the RA; the even unknown numbers correspond to the declination of the sources.

## 4  Summary

Our analysis revealed substantial differences between rotation parameters between CRF realizations (radio source position catalogues) computed without accounting for correlation information and using full covariance matrices of individual catalogues. The difference in the rotation angles may exceed 20 μas (see Table 4). Thus our findings confirm result of Jacobs et al. (2010) based on different catalogues and comparison scheme. Therefore, it is necessary to account for the full covariance information during comparison, combination and analyses of modern CRF solutions.

On the other hand, accounting for RA/DE correlations only, the differences in rotation parameters are found to be at a level below 1σ, i.e. practically insignificant (cf. the first two lines in Tables 2-4).

It should be noted that accounting for a full correlation matrix may be essential not only for definition of mutual orientation, but also for decomposition of the coordinate differences by orthogonal functions. If it is the case, the full correlation information should be accounted for during calculation of a combined catalogue as proposed by Sokolova and Malkin (2007). Corresponding investigations are underway.

Indeed, the results obtained in this study can be applied also to determination of orientation parameters between other reference frames, such as TRF solutions from space geodesy techniques or optical source position catalogues.

## 5 Acknowledgements

The authors are grateful to all the VLBI data analysts, who made their catalogues available for this study either via public access or via personal contact. We thank the anonymous referees for careful reading of the manuscript and useful comments and suggestions.

## 6 References

Arias EF, Lestrade J-F, Feissel M (1988) Comparison of VLBI celestial reference frames. Astron Astr 199:357–363

Feissel M, Mignard F (1998) The adoption of ICRS on 1 January 1998: meaning and consequences. Astron Astr 331:L33–L36

Jacobs CS, Heflin MB, Lanyi GE, Sovers OJ, Steppe JA (2010) Rotational alignment altered by source position correlations. In: Behrend D, Baver KD (eds) Proceedings of IVS 2010 General Meeting, pp 305–309.

Ma C, Arias EF, Eubanks TM, Fey AL, Gontier A-M, Jacobs CS, Sovers OJ, Archinal BA, Charlot P (1998) The international celestial reference frame as realized by very long baseline interferometry. Astron J 116:516–546

Ma C, Arias EF, Bianco G, Boboltz DA, Bolotin SL, Charlot P, Engelhardt G, Fey AL, Gaume RA, Gontier AM, Heinkelmann R, Jacobs CS, Kurdubov S, Lambert SB, Malkin ZM, Nothnagel A, Petrov L, Skurikhina E, Sokolova JR, Souchay J, Sovers OJ, Tesmer V, Titov OA, Wang G, Zharov VE, Barache C, Boeckmann S, Collioud A, Gipson JM, Gordon D, Lytvyn SO, MacMillan DS, Ojha R (2009) The second realization of the international celestial reference frame by very long baseline interferometry. In: Fey AL, Gordon D, Jacobs CS (eds) IERS Technical Note No. 35, Verlag des Bundesamts für Kartographie und Geodäsie, Frankfurt am Main

Malkin Z (2013) A new approach to the assessment of stochastic errors of radio source position catalogues. Astron Astr 558:A29. doi: 10.1051/0004-6361/201322334

Sokolova J, Malkin Z (2007) On comparison and combination of catalogues of radio source positions. Astron Astr 474:665–670. doi: 10.1051/0004-6361:20077450